\documentstyle[epsfig]{aipproc}

\begin{document}
\title{Hadronic Decays of Beauty and Charm from CLEO}
\author{Jorge L. Rodriguez \\ 
(CLEO Collaboration)}
\address{Department of Physics and Astronomy University of Hawaii\\
2505 Correa Road Watanabe Hall 227\\
Honolulu, Hawaii 96822}

{\hskip 11.8cm uh-511-920-98}
\maketitle

\def\etal{{\it et al.}}
\def\ie{{\em ie,}}
\def\bmass{$M_{BC}\ $}
\def\ediff{$\Delta \rm{E}\ $}
\def\bbar{$B\bar{B}\ $}


\def\REF{Ref.}
\def\equ#1{Equation (#1)}
\def\tbl{Table}
\def\fig{Figure}
\def\cleoii{CLEOII\ }
\def\cleosvx{CLEOII/SVX\ }

\def\ddblstrpi{B^-\rightarrow D_J^{0}\pi^{-}}
\def\dstrdstrs{B^0\rightarrow D^{*-}D_s^{*+}}
\def\dstrdstr{B^0\rightarrow D^{*-}D^{*+}}
\def\dstrrho{B\rightarrow D^*\rho}
\def\dstrprho{\bar{B}^0\rightarrow D^{*+}\rho^-}
\def\dstrzrho{B^-\rightarrow D^{*0}\rho^-}
\def\thdst{\theta_{D^*}}
\def\thrho{\theta_{\rho}}
\def\Hp{H_+}
\def\Hm{H_-}
\def\H0{H_0}
\def\dsp{D^{*\prime +}}
\def\Kpiws{D^0\rightarrow K^+\pi^-}
\def\Kpirs{D^0\rightarrow K^-\pi^+}

\def\europhy{{\it Eur. Phys. J.,\ }}
\def\zphy{{\it Z. Phys.,\ }}
\def\nim{{\it Nucl. Inst. Meth.,\ }}
\def\pr{{\it Phys. Rev.,\ }}
\def\prd{{\it Phys. Rev. D,\ }}
\def\prl{{\it Phys. Rev. Lett.,\ }}
\def\npb{{\it Nucl. Phys. B,\ }}
\def\pl{{\it Phys. Lett.,\ }}
\def\plb{{\it Phys. Lett. B,\ }}

\begin{abstract}
A selection of recent results on hadronic charm and beauty decays from
the CLEO experiment are presented. We report preliminary evidence
for the existence of final state interactions in $B$ decays and the first
observation of the decay $B^0\rightarrow D^{*+}D^{*-}$ with a
branching fraction of $(6.2^{+4.0}_{-2.9}\pm 1.0) \times 10^{-4}$. We
also present preliminary results on the first observation of the
broad, $J^P=1^+$, charmed meson resonance with a mass of
$m_{D_1(j=1/2)^0} = 2.461^{+0.041}_{-0.034}\pm 0.010\pm 0.032$ GeV and a
width of $\Gamma = 290^{+101}_{-79}\pm 26 \pm 36$ MeV and branching fraction
measurements of the $B^-\rightarrow D_J^0\pi^-$\footnote{Unless
otherwise indicated complex conjugate states are implied throughout
this paper} decay. Finally, we report on our search for the radial
excitation of a spin 1 charmed meson, the $D^{*\prime +}$, and on
an improved measurement of the ratio of decay rates
$\Gamma(D^0\rightarrow K^+\pi^-)/\Gamma(D^0\rightarrow
K^-\pi^+)$. 
\end{abstract}

\section*{Introduction}
The CLEO experiment has provided important contributions to our
understanding of hadronic decays of the beauty and charm systems since
it began taking data in the early 1980s. The wealth of results is due primarily
to the large data samples collected over the years and the excellent
tracking, energy resolution and reasonably good particle ID of the
CLEO series of detectors. In this paper we present five analyses on
hadronic decays of charm and bottom mesons. 

CLEO is currently analyzing data from two separate runs taken with
different detector configurations. The first run ended in the Summer of
1995 and includes a total luminosity of 3.1 $fb^{-1}$ on and 1.6
$fb^{-1}$ taken 60 MeV below the $\Upsilon(4S)$. Given the \bbar cross
section this sample corresponds to 3.1 million \bbar pairs. The
configuration of the detector during the first run is described in
detail in \REF~\cite{jrod:CLEOII}. This dataset will be
referred to as the \cleoii dataset hereafter. At the end of the
\cleoii run the detector was significantly improved with the replacement
of the inner straw tube drift chambers by a 3 layer silicon vertex
(SVX) detector\cite{jrod:CLEOSVX}.  In addition, the argon-ethane gas
in the drift chambers was replaced with a helium-propane mixture which
improved both particle ID and the momentum resolution in the drift
chambers. Finally, the track fitting software was updated to one based
on the Kalman filtering algorithm. These improvements in tracking and
particle ID, while featured in only two of the analysis presented here
will be become increasingly important in future analyses. The data
collected and reconstructed, with the \cleoii upgrade, (CLEOII/SVX),
consists of 2.5 $fb^{-1}$ on and 1.3 $fb^{-1}$ off resonance. The data
run for \cleosvx will be completed at the end of 1998.

\subsection*{First Observation of \mbox{\boldmath $\dstrdstr$}}
The Cabbibo suppressed decay $\dstrdstr$ is a potentially interesting
CP violation mode, whose rate is expected to be comparable to the gold
plated CP $B^0\rightarrow J/\Psi K_s$ decay. Since the $D^{*+}D^{*-}$
final state can be obtained from either $B^0$ or a $\bar{B}^0$ this
decay mode mode can be used to extract $\sin 2\beta$ through
$B^0\bar{B^0}$ mixing.  The amplitude for this decay is dominated by
the external tree diagram and we can estimate its rate by comparison to
the measured $\dstrdstrs$ rate, after taking into account the
appropriate ratio of decay constants and CKM matrix elements. 
While the expected rate is of order 0.1 \%, the rather large number of
particles, six in the lowest multiplicity mode, in the decay chain 
significantly reduces the expected yield.

CLEO has performed a search for this mode by examining all of the
currently available data collected on the $\Upsilon(4S)$
\cite{jrod:dstrdstr}. This includes the complete \cleoii (3.1
$fb^{-1}$) and the available portion of \cleosvx data (2.5
$fb^{-1}$). The decay chain is fully
reconstructed cutting on kinematic variables to reduce backgrounds. In
this analysis only three of the possible four combinations of the 
$D^{*+}D^{-*}$ were used. The decay mode with two soft $\pi^0$, the
$B^0\rightarrow (D^+\pi^0)(D^-\pi^0)$ decays was not used due to background
considerations. For events in the \cleosvx
sample an additional requirement was imposed to take advantage of the
better position resolution obtained from the SVX. 


\begin{figure}[t!] 
\centerline{\epsfig{file=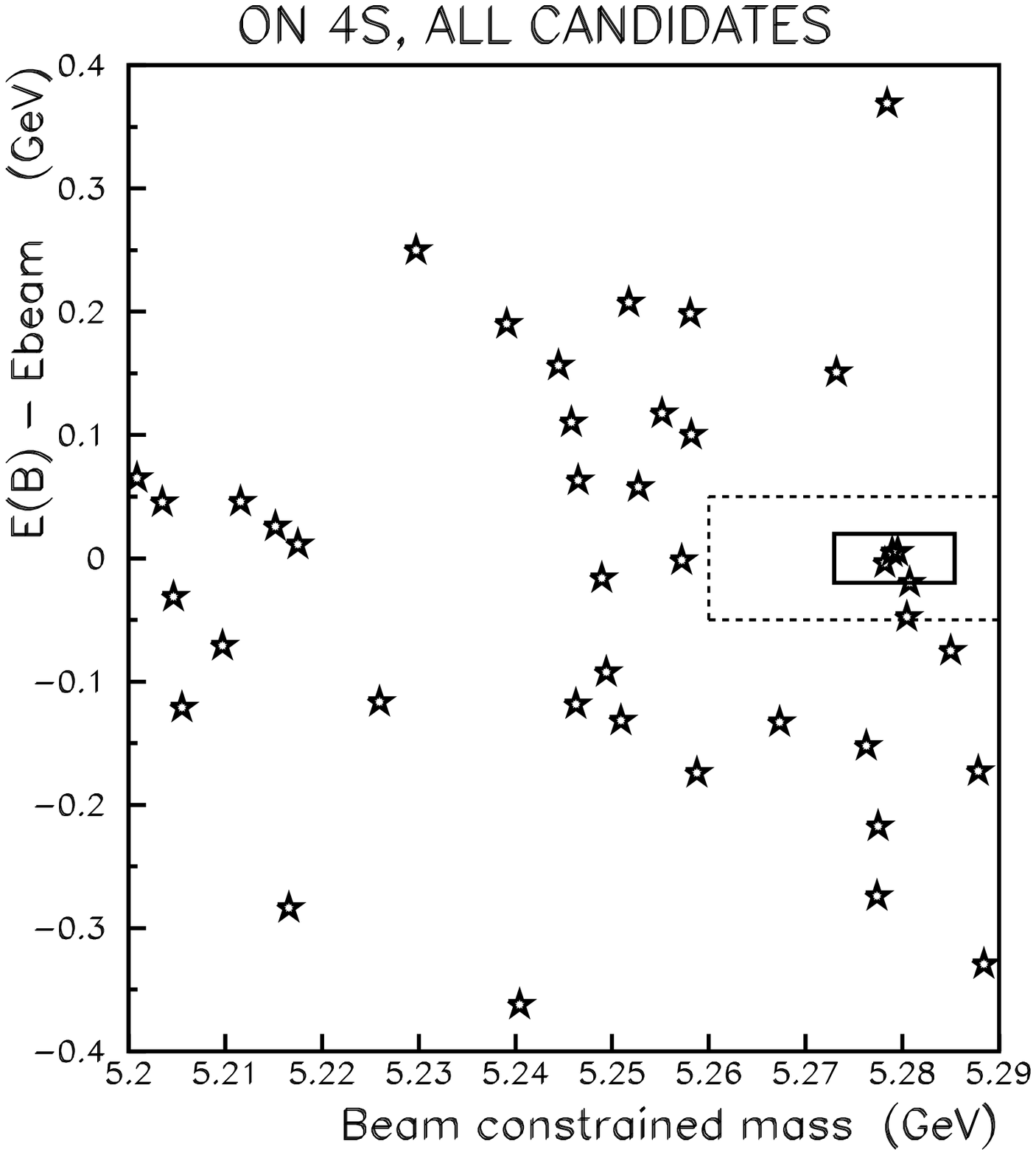,height=3.0in}\epsfig{file=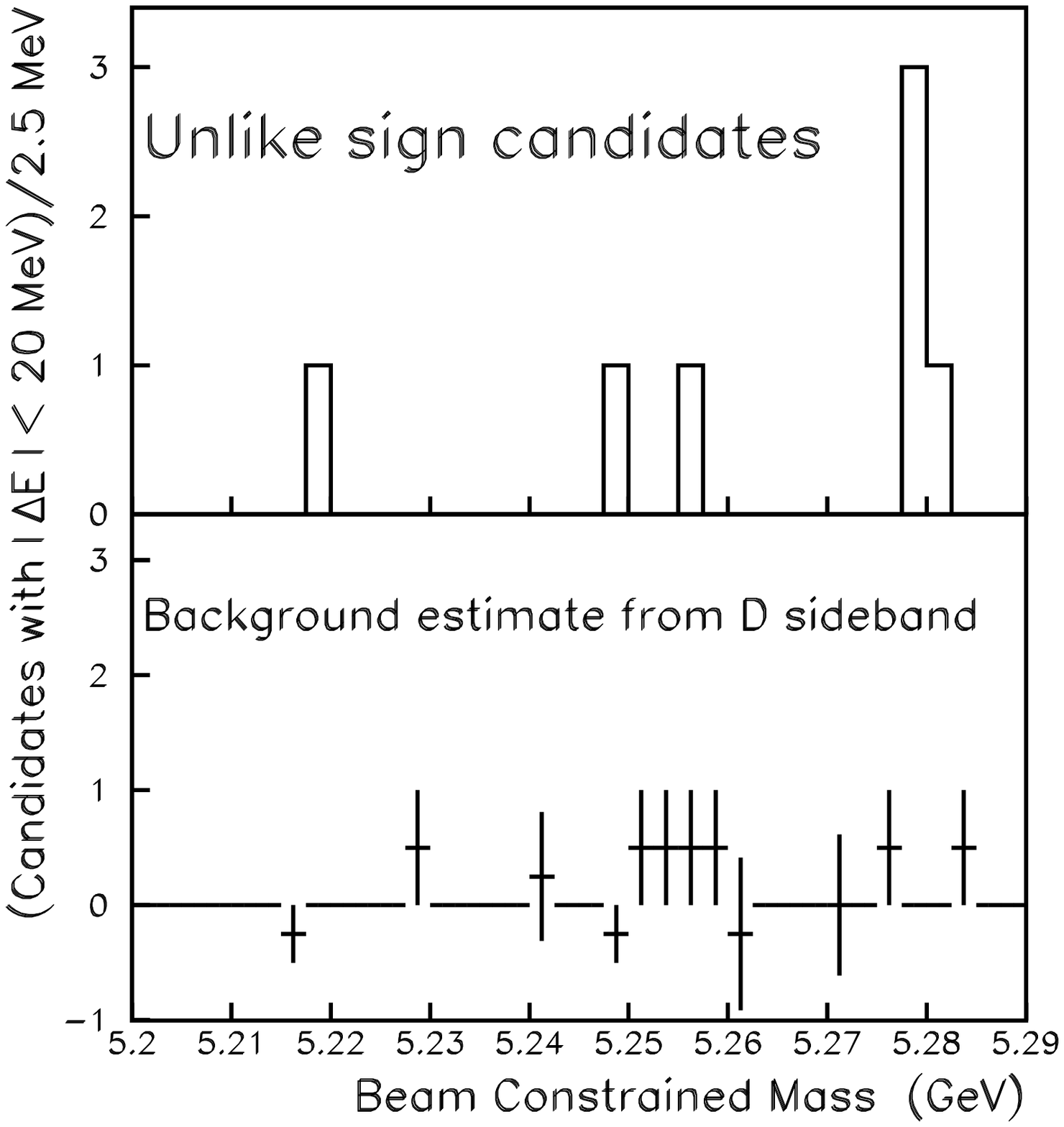,height=3.in}}
\vskip .3 truein
\caption{Scatter plot of on 4S events in \ediff and \bmass
(left). Beam-constrained mass distribution for events that lie within
20 MeV of 0.0 in \ediff (right-top). The plot (right-bottom) shows the
backgrounds from the $D$ mass sidebands.}
\label{jrod:fig:dstdstsig}
\end{figure}

The observables used to extract the signal were the beam-constrained
mass ($M_{BC}$) and the difference between the reconstructed energy and
the beam energy ($\Delta E$). At CLEO, the $B$s are produced nearly at
rest so \ediff, for real events, is peaked at zero while backgrounds
from other decays peak one or more pion mass away from zero.  The
beam-constrained mass variable is just the usual invariant mass with
the beam energy substituted for the measured energy. The resolution of
\bmass is significantly better than the invariant mass, by about an
order of magnitude, due to the small spread of the beam energy. In
\fig~\ref{jrod:fig:dstdstsig} we show the scatter plots of the
on-$\Upsilon(4S)$ distributions for events that pass all of event
selection criteria. The solid rectangle in \fig~\ref{jrod:fig:dstdstsig} 
(left) is the signal region. A total of 4 events were observed.

To estimate the backgrounds that enter into the signal region, two
independent methods were used. The first estimate is based on events
in the \ediff vs. \bmass sideband indicated by the region outside
dashed rectangle in \fig~\ref{jrod:fig:dstdstsig} (left). An estimate
of $0.26\pm 0.04$ events is determined from this sideband estimate. A
second estimate is obtained by adding contributions from continuum,
\bbar (other kinematically similar $B$ decays that can fake the
signal) and random combination that are reconstructed as signal. Each
of these contributions were modeled by off-resonance data, Monte
Carlo, and/or $D$ mass sidebands. This estimate predicts a background
of $0.37\pm 0.05$ events.

The branching fraction measured given the four observed events is:
\begin{equation}
{\cal B}(\dstrdstr) = (6.2^{+4.0}_{-2.9}\pm 1.0)\times 10^{-4}.
\end{equation}
\noindent This value is determined from an unbinned likelihood fit
using the larger of the two background estimates.  This value is
consistent with the expected rate of 0.1\% given the measured
branching fraction of the $\dstrdstrs$, our knowledge of decay
constants and the CKM matrix elements~\cite{jrod:dstrdstr}.


\subsection*{Angular Distributions in \mbox{\boldmath $\dstrrho$}}
A full partial wave analysis of the decays $\dstrzrho$ and $\dstrprho$ has
been performed using the entire \cleoii data sample. These decays
proceed primarily through tree-level $b\rightarrow c$ $W$ emission
and, to first order, the amplitude for the decays
are independent of a CKM phase. The absence of a weak phase
suggests a clean model to study the effects of final state
interactions (FSI) in hadronic $B$ decays. The full partial wave
decomposition, with its own phases, provides us with a way to 
determine the strong phases through an analysis of the angular distribution 
of the final states.


In order to extract information on the strong phases we first need to
express the differential decay rate in terms of complex amplitudes
and helicity angles. In this analysis we use the helicity basis
expressed in three components; two, the $H_{\pm}$, represent the
transverse components and one, the $H_0$ describes the longitudinal
component. Squaring and factoring the amplitude gives the
differential decay rate in terms of the helicity amplitudes and the
helicity angles $\theta_{\rho},\theta_{D^*}$ and $\chi$. The form of
the expression is,

\begin{eqnarray}
{d\Gamma\over d\cos\thdst d\cos\thrho d\chi} =
4|\H0|^2\cos^2\thdst\cos^2\thrho + \left(|\Hm|^2+|\Hp|^2\right)\sin^2\thdst\sin^2\thrho\nonumber\\
+ 2\left[Re(\Hp\Hm^*)\cos 2\chi - Im(\Hp\Hm^*)\sin
2\chi\right]\sin^2\thdst\sin^2\thrho \nonumber \\
+\left[Re(\Hp\H0^*+\Hm\H0^*)\cos\chi - Im(\Hp\H0^*-\Hm\H0^*)\sin\chi\right]\sin 2\thdst\sin 2\thrho.
\label{jrod:eq:angdist}
\end{eqnarray}

The two helicity angles are defined in the rest frame of the decay as
the angle between one of the daughters and the direction of the parent
in the rest frame of the $B$. The angle $\chi$ is the angle between
the two decay planes and is related to the azimuthal direction of the
helicity axis by $\chi=\phi_{D^*}-\phi_{\rho}$. In the amplitude, the
strong phase information is contained in the terms with the imaginary
parts; in \equ{\ref{jrod:eq:angdist}}, no FSI implies that
either $Im(\Hp\H0^*-\Hm\H0^*)$ and $Im(\Hp\Hm^*)$ are zero,
or conversely, that all amplitudes are relatively real\cite{jrod:KMP92}.

All events are required to pass a series of
selection criteria to fully reconstruct the decay chain of the $B$
using three decay modes of the $D^0$, the $K\pi,K\pi\pi^0$ and $K3\pi$,
and the dominant decay modes of the $D^*$\cite{jrod:Dstrpol-conf}. Two
methods are used to extract the phase information in
\equ{\ref{jrod:eq:angdist}}: a moments analysis\cite{jrod:DDF98} in
which the components of each term in \equ{\ref{jrod:eq:angdist}} are
extracted and a direct determination of the magnitude and phases of
the helicity amplitudes from a
three dimensional (3D) unbinned maximum likelihood fit of the data to
the functional form in \equ{\ref{jrod:eq:angdist}}.

In the unbinned likelihood fit and moments analysis the likelihood
function includes terms for the signal and background contributions,
factorizing each term into an angular part and a mass part. The mass
part characterizes the $\rho$ invariant mass with a relativistic
Breit-Wigner and the beam-constrained mass with a Gaussian
function. The angular part is modeled by \equ{\ref{jrod:eq:angdist}}.
To minimize the number of free parameters the fit is first performed
ignoring the angular part and the parameters of the mass function are
extracted. In the second step, the fit is redone including the angular
function and fixing the mass parameters to values extracted from the
first fit.  The parameters in the angular part of the likelihood
function are the phases and magnitudes of the transverse helicity
amplitudes relative to the longitudinal component which is set to $H_0
= 1$ and $\delta_0 = 0$ in the fit. The amplitudes are then rescaled
to the satisfy the normalization condition $|H_0|^2+|H_-|^2|H_+|^2 = 1$
The results of the likelihood fit
for the strong phases and amplitudes are given in \tbl~\ref{jrod:tbl:pol1}. 
The coefficients of \equ{\ref{jrod:eq:angdist}} have also been
determined from the fit and a comparison made with the values obtained
from the moments analysis. We find that the results are consistent
with each other within statistical errors (see \tbl~\ref{jrod:tbl:pol2}). 
Our values for the phases in \tbl~\ref{jrod:tbl:pol1} suggest
non-trivial strong phases in both the $\dstrzrho$ and $\dstrprho$
modes, however the statistical size of our sample does not provide us
with an independent confirmation of the results in a 1D fit of the
data to the $\sin\chi$ or $\sin2\chi$ distributions. 

The results of the fit are also used to test the factorization
hypothesis by comparing the polarization of the $\dstrprho$ decay
with the polarization in the semi-leptonic $\bar{B}^0\to D^{*+}l^-\bar{\nu}$ 
at the appropriate $q^2$ scale. The predicted values for the longitudinal
polarization in the semi-leptonic decay at $q^2=m_\rho^2$, range
from 85\% to 88\% and the recent CLEO results is $91.4\pm 15.2\pm 8.9
\%$\cite{jrod:Dstrpol-conf}. The longitudinal polarization from the fit is 
$87.8 \pm 3.4\pm 4.0 $ \% consistent, within errors, with both the
theoretical predictions and the semi-leptonic measurement.

\begin{table}[h!]
\caption{Phases extracted from the unbinned likelihood fit}
\label{jrod:tbl:pol1}
\begin{tabular}{|l||ll|}
\noalign{\vspace{-8pt}} \multicolumn{1}{|c||}{Parameter}  &
\multicolumn{1}{c}{$\dstrprho$} & \multicolumn{1}{c|}{$\dstrzrho$} \\ \cline{1-3}    
$\delta_-$ & $0.19\pm 0.23\pm 0.14$    & $1.13\pm 0.27\pm 0.17$     \\
$\delta_+$ & $1.47\pm 0.37\pm 0.32$    & $0.95\pm 0.31\pm 0.19$     \\ 
$|H_-|$    & $0.317\pm 0.052\pm 0.013$ & $0.283\pm 0.068\pm 0.039$  \\
$|H_+|$    & $0.152\pm 0.058\pm 0.037$ & $0.228\pm 0.069\pm 0.036$  \\ 
\end{tabular}
\end{table}

\begin{table}[b]
\caption{Coefficients in Equation (3) extracted from the likelihood fit}
\label{jrod:tbl:pol2}
\begin{tabular}{|l||ll||ll|}
\noalign{\vspace{-8pt}}	& \multicolumn{2}{|c||}{$\dstrprho$} & \multicolumn{2}{|c|}{$\dstrprho$} \\ \cline{2-5}
\multicolumn{1}{|c||}{Coefficient}	& \multicolumn{1}{|c}{From L.L. fit\rule[-4pt]{0pt}{17pt}} &\multicolumn{1}{|c||}{From moments} &\multicolumn{1}{|c}{From L.L. fit\rule[-4pt]{0pt}{17pt}} &\multicolumn{1}{|c|}{From moments}\\
\hline
$H_0^2$                 & $0.856$          & $0.751\pm 0.073$ & $0.859$          & $0.626\pm 0.074$ \\
$H_+^2+H_-^2$           & $0.140\pm 0.040$ & $0.159\pm 0.034$ & $0.143\pm 0.060$ & $0.168\pm 0.036$ \\
$Im(H_-H_0^*-H_+H_0^*)$ & $0.110\pm 0.074$ & $0.042\pm 0.103$ &$-0.071\pm 0.109$ &$-0.145\pm 0.101$ \\
$Re(H_-H_0^*+H_+H_0^*)$ & $0.341\pm 0.088$ & $0.352\pm 0.104$ & $0.250\pm 0.105$ & $0.193\pm 0.109$ \\
$Im(H_+H_-^*)$          & $0.053\pm 0.021$ & $0.057\pm 0.024$ &$-0.011\pm 0.032$ & $0.002\pm 0.027$ \\
$Re(H_+H_-^*)$          & $0.023\pm 0.024$ & $0.018\pm 0.023$ & $0.068\pm 0.029$ & $0.043\pm 0.025$ 
\end{tabular}
\end{table}





\begin{figure}[t!] \centerline{\epsfig{file=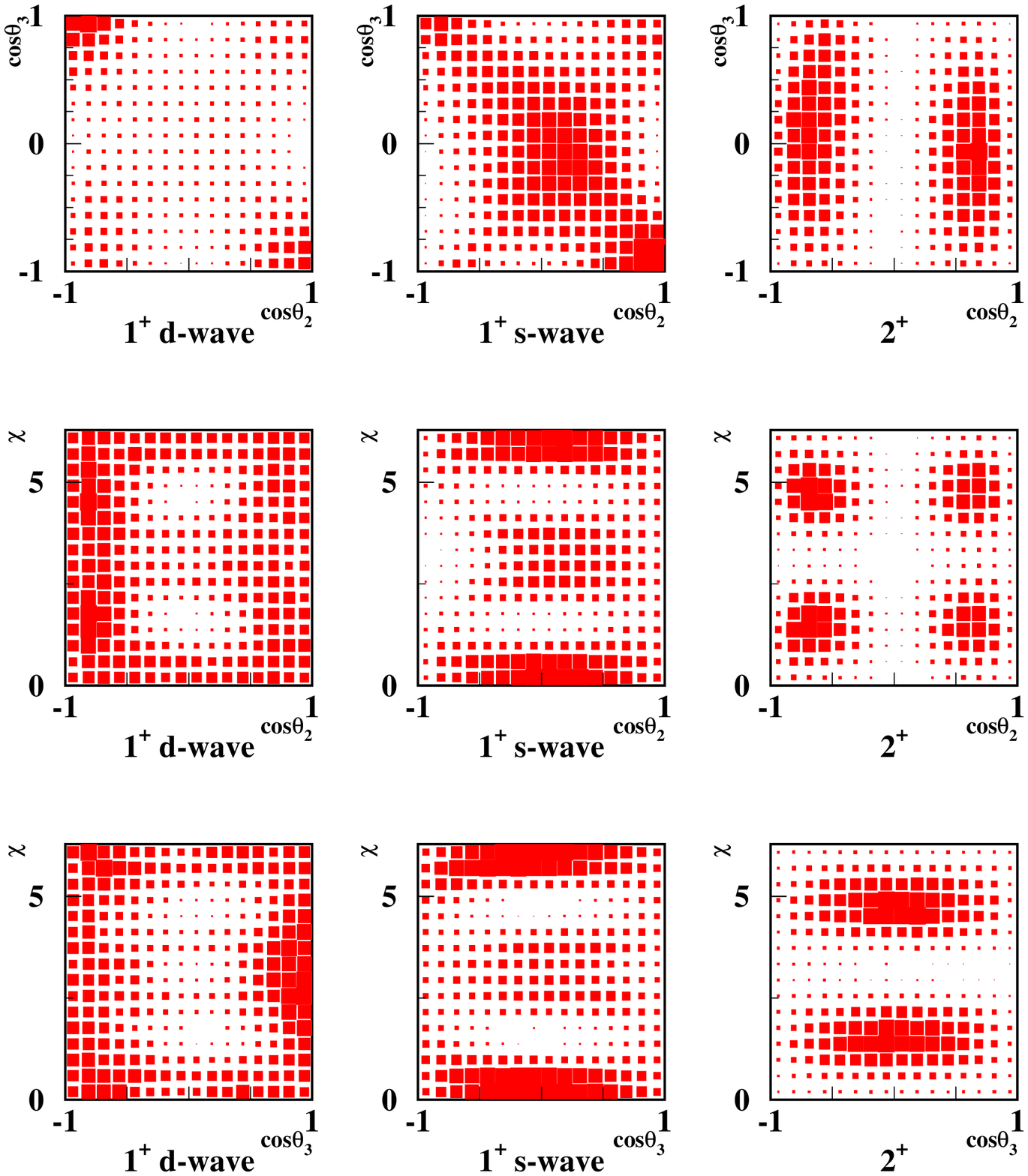,height=3.2in}\epsfig{file=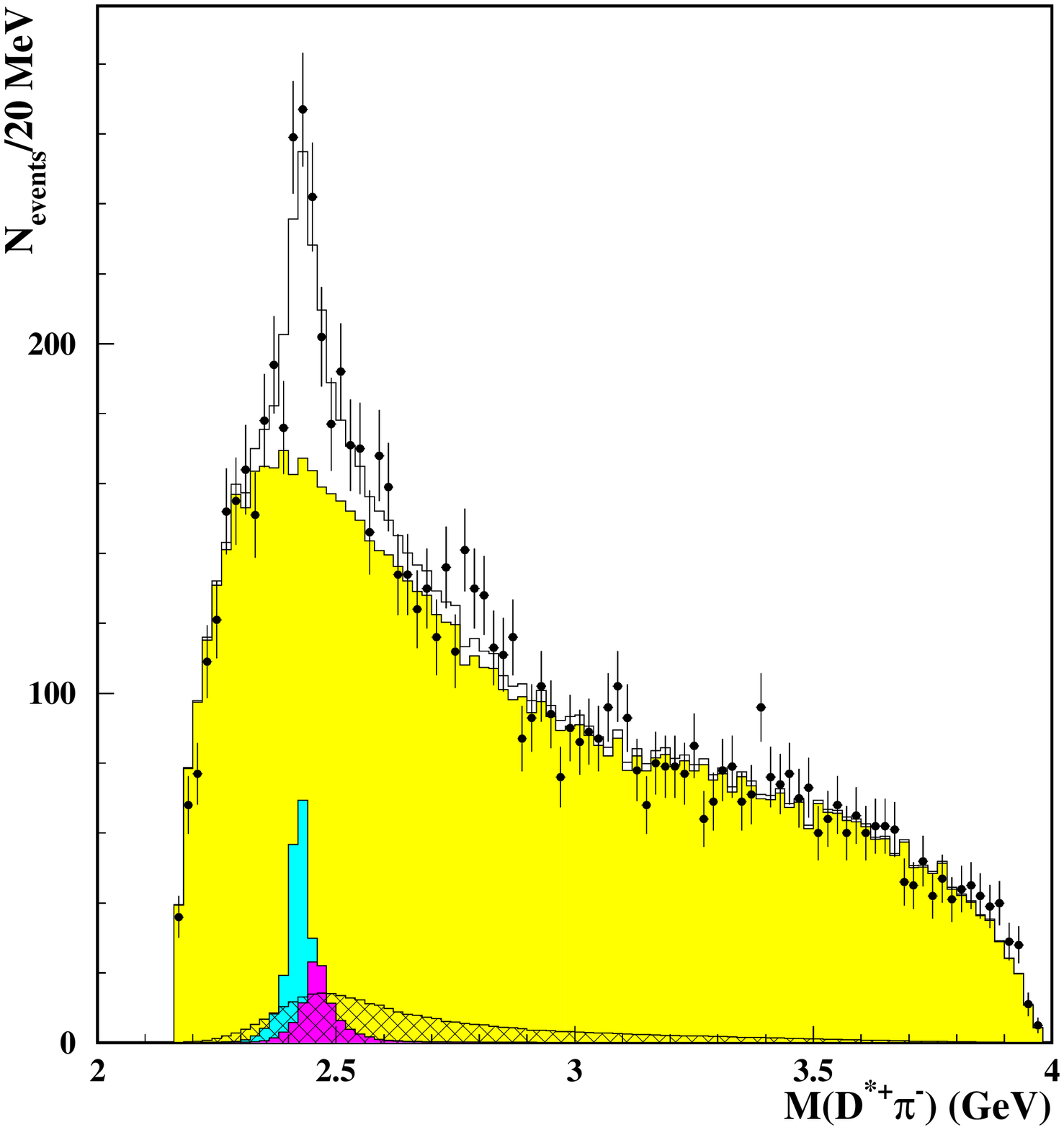,height=3.3in}}
\caption{The reconstructed 2D angular distributions for the decays of
the $B^-$ meson to each of the three $D_J^0$ resonances (left). The
$D^{*+}\pi^-$ invariant mass distribution for on 4S data with the
4D-MLF projections from each of the three $D_J^0$ candidates, the two
narrow $1^+$ (light) and the $2^+$ (darker) resonance and the broad
$1^+$ (cross-hatched) plus the total background are superimposed on
the plot (right).}
\label{jrod:fig:ddblstar}
\end{figure}

\subsection*{Measurements in \mbox{\boldmath $B^-\rightarrow D^0_J\pi^-$}}
The $L=1, n=1$ charmed mesons are the $P$ wave orbital excitations in which
four spin-orbit configurations are possible. In the heavy quark limit,
these combinations can be identified by the $j_l$ quantum number which
couples the spin of the light quark with the orbital angular
momentum. These states form two doublets: a $j_l=3/2$ and a $j_l=1/2$
which, from conservation of parity and angular momentum, decay via
either $S$ or $D$ waves. In the heavy quark limit, the $j_l=1/2$
decays only via $S$ wave while the $j_l=3/2$ can decay only via $D$
wave. The only $L=1$ states so far observed have been the narrow
$L=1,n=1$ resonances, the $D_1(2420)$ and the $D_2^*(2460)$ with
widths of order 20 MeV. These states have been assigned to the
$j_l=3/2$ by observing the angular distributions of the decay products
and measurements of the ratio of $D^*\pi$ to $D\pi$ decays in
continuum production where the $D_J$ are unpolarized.

A full partial wave analysis of the decay $B^-\rightarrow D^0_J\pi_1^-,
D^0_J\rightarrow D^{*+}\pi_2^-,D^{*+}\rightarrow D^0\pi_3^+$ has been
performed to measure the product branching fraction of the ${\cal
B}(B)\times{\cal B}(D_J)$ decays and the properties, the mass and
width, of the broad $D_1(j=1/2)$ resonance. The measurements are
extracted from a 4D unbinned maximum likelihood fit
(4D-MLF) to the data, where the independent variables are the three
helicity angles and the invariant mass of the $D_J^0$ resonance. 

An important point in this analysis is the fact that the $D_J$ is
completely polarized since it is the decay product of a
pseudo-scalar decay to a vector plus another pseudo-scalar. Knowing
the initial polarization of the $D_J$ and the fact that angular
momentum and parity are both conserved provides us
with a clear picture, in the heavy quark limit, of the angular
decay distribution in final states that first decay through one of the
three intermediate $D^0_J$ resonances. In other words, we can
distinguish among the three possible $L=1$ states, those that decay to
a $D^{*+}\pi^-$, not only by using the invariant mass but also by
examining the full angular distribution of the final state.


A partial reconstruction technique selects the events from among the
entire \cleoii dataset. These events are used in the fit to the 4D
maximum likelihood function. In the partial reconstruction method the
entire decay is reconstructed, up to a quadratic ambiguity, from the
4-momenta of the three pions ($\pi_1,\pi_2,\pi_3$) in the decay chain
and imposing energy-momentum conservation at each decay
vertex\cite{jrod:part_recon}. This method improves statistics by about
an order of magnitude over the usual full reconstruction technique
since it eliminates the explicit reconstruction of the charmed
meson. A trade off to the gain in statistics comes from the increased
complexity of the analysis and higher levels of backgrounds. These
backgrounds are however, modeled by using the 1.6 $fb^{-1}$ of
off-resonance data and Monte Carlo simulations. In
\fig~\ref{jrod:fig:ddblstar} (left) we show the $D^{*+}\pi^-$
invariant mass distribution taken from events that pass the selection
criteria described in \REF~\cite{jrod:ddblstarconf} in the
on-resonance \cleoii dataset. Superimposed on the plot are the
$D^{*+}\pi^-$ invariant mass projection from the 4D-MLF for each of
$D_J^0$ candidates plus the total background contribution from various
sources; continuum and other \bbar decays with similar kinematics. The
ability of the angular information to distinguish between the three
possible resonances is illustrated in \fig~\ref{jrod:fig:ddblstar}
(right) where Monte Carlo simulations of the $B^-$ decays to each of
the three $L=1$ $D_J$ resonances are shown.

The 4D likelihood function used in the fit includes terms for the
angular distribution, mass amplitudes of the resonances, strong phase
shifts and parameters that allow for mixing between the two $1^+$
states. It also allows for detector smearing and acceptance. The 
functional form of the amplitude is 
\begin{eqnarray}
{\cal A}_{B\rightarrow D^{*+}\pi^-\pi^-} =&
\alpha_{nr}e^{i\delta_{nr}} + \alpha_2 A_2 a_2 e^{i\delta_2}
+ \alpha_{1n}A_{1n}\left(a_{1d}\cos\beta + a_{1s}\cos\beta
e^{i\phi}\right)\nonumber \\
&+ \alpha_{1b}A_{1b}\left(a_{1s}\cos\beta - a_{1d}\cos\beta e^{i\phi}\right)e^{i\delta_1}
\label{jrod:eq:grand-amp}
\end{eqnarray}
\noindent where the $\alpha_i$ allows for different contributions from
the various resonant and non-resonant $D^{*+}\pi^-\pi^-$ components,
the $A_i$ are Breit-Wigner amplitudes, and the $a_i$ are the angular
($D^j_{m,m^\prime}$) amplitudes. The mixing between the narrow and
broad $1^+$ resonances is currently described by the mixing angles
$\beta$ and $\phi$ and the strong phases for the resonant and
non-resonant components are included via the $\delta$ parameters. The
$1n$,$2$,$1b$, and $nr$ subscripts refer to the narrow $1^+,j_l=3/2$ and
$2^+,j_l=3/2$ resonances, the broad $1^+,j_l=1/2$ resonance and the non-resonant
component respectively. This
parameterization is not unique and an alternative parameterization has
been used as a systematic check. The variation in the results due to
the alternative parameterization is quoted as an additional systematic
error. The fit is performed with the mass and width of the
relativistic Breit-Wigners for the narrow $2^+$ and the $1^+$ states
fixed to their measured values\cite{jrod:PDG}. The normalization of
each of the three resonant and the non-resonant components plus the
mass and width of the broad $1^+$ state are allowed to float in the
fit. From the fit we extract the invariant mass and width of the broad
$1^+$ resonance to be:

\begin{eqnarray}
M_{D_1(j=1/2)^0}& =& 2.461^{+0.041}_{-0.034}\pm 0.010\pm 0.032 \quad {\rm
GeV} \nonumber \\
\Gamma_{D_1(j=1/2)^0} &=& 290^{+101}_{-79}\pm 26 \pm 36 \quad {\rm MeV}
\end{eqnarray}
\noindent where the first error is the statistical uncertainty, the
third is the uncertainty from the parameterization of the amplitude and
the second is the systematic uncertainty from all other sources.  From
the 4D fit we also extract the product branching ratios of the $B^-$
to each of the three $D_J^0$ plus a single pion. These results are
given in \tbl~\ref{jrod:tbl:ddblstrpi} together with the yields and
the $B^-$ branching fractions using the $D^0_J$ branching fractions
from isospin symmetry. The second systematic error in
\tbl~\ref{jrod:tbl:ddblstrpi} represents the uncertainty in the
parameterization of the grand-amplitude. These results are consistent
with the values obtained earlier using a simpler 2D-MLF where not all
of the angular information was used\cite{jrod:ddblstarconf}. These
branching fraction measurements, however, disagree with theoretical
expectations from heavy quark effective theory which predict the rates
to be about three times smaller than the our
results\cite{jrod:N97}. Our preliminary results on the mass and width
of the broad $1^+$ charmed meson are in agreement with the quark
model\cite{jrod:GK91}.

\begin{table}[b!]
\caption{Results of the 4D maximum likelihood fit to the decay $\ddblstrpi$}
\label{jrod:tbl:ddblstrpi}
\begin{tabular}{|l||l|l|l|} 
\noalign{\vspace{-7pt}} \multicolumn{1}{|c||}{$B$ Decay Mode}  &
\multicolumn{1}{c|}{\ \ Event Yield \ } & 
\multicolumn{1}{c|}{${\cal B}(B^-\rightarrow D^0_J\pi^-)\tablenote{We
use the $D_J$ branching fraction assumed in much of the literature
${\cal B}(D_1^0\rightarrow D^{*+}\pi^-)=2/3$ and  ${\cal B}(D_2^{*0}\rightarrow D^{*+}\pi^-)=0.2$}\times 10^{-3}$}& 
\multicolumn{1}{c|}{$\ {\cal B}(B^-)\cdot {\cal B}(D^0_J)\times 10^{-4} $} \\ \cline{1-4}
$D_1^0(j_l=1/2)\pi^-$& $237.1\pm 42$& $1.59\pm 0.29\pm 0.26\pm 0.03\pm 0.035$       & $10.6\pm 1.9\pm 1.7\pm 2.3 $    \\
$D_1(2420)^0\pi^-$   & $420.0\pm 41$& $1.04^{+0.27}_{-0.21}\pm 0.17\pm 0.02\pm 0.07$& $6.9^{+1.8}_{-1.4}\pm 1.1\pm 0.4$\\ 
$D_2^*(2460)^0\pi^-$ & $109.5\pm 26$& $1.55\pm 0.42\pm 0.23\pm 0.03\pm 0.14$        & $3.1\pm 0.84\pm 0.45\pm 0.28$    \\
$D^{*+}\pi^-\pi^- $  & $160.0\pm 61$& $9.7\pm 3.6\pm 1.5\pm 1.9$
&\\ \hline
Total		     &              & $29.2\pm 4.5\pm 3.7\pm 3.1$                   &\\                                  
\end{tabular}
\end{table}

\subsection*{Search for First Radial Excitations in Charmed Meson}
In 1997 the DELPHI collaboration claimed evidence for the $1^{st}$ radially
excited charmed meson $\dsp$\cite{jrod:DELPHI}. They found an excess
of $66\pm 14$ events in their sample of reconstructed $D^{*+}\pi^-\pi^+$ 
with a mass of $2637\pm 2\pm 6$ MeV and a small width. The assignment of the
quantum numbers was based primarily on the mass measurement which is 
consistent with theoretical expectations for a $\dsp$\cite{jrod:GI85}. 
The width of the bump is approximately equal to the detector
resolution so DELPHI sets an upper bound on the decay width of the $\dsp$ to
be $< 15$ MeV at the 95\% confidence level. The
OPAL experiment has also performed a search for the $\dsp$ in the same
final state and in the DELPHI mass window using a similar analysis
procedure. They however, found no excess and set an upper limit on 
$\dsp$ production of $f_{Z^0\rightarrow \dsp }\times {\cal B}(\dsp \rightarrow
D^{*+}\pi^-\pi^+)<2.1\times 10^{-3}$ \@ 95\% C.L.\cite{jrod:OPAL98}.
Both experiments collect data at the $Z^0$ mass so $\dsp$ production
is from the $c\bar{c}$ and/or $b\bar{b}$ continuum. Both
experiments also estimate that about half of their candidates are from
$c\bar{c}$ production.

\begin{figure}[t!] 
\centerline{\epsfig{file=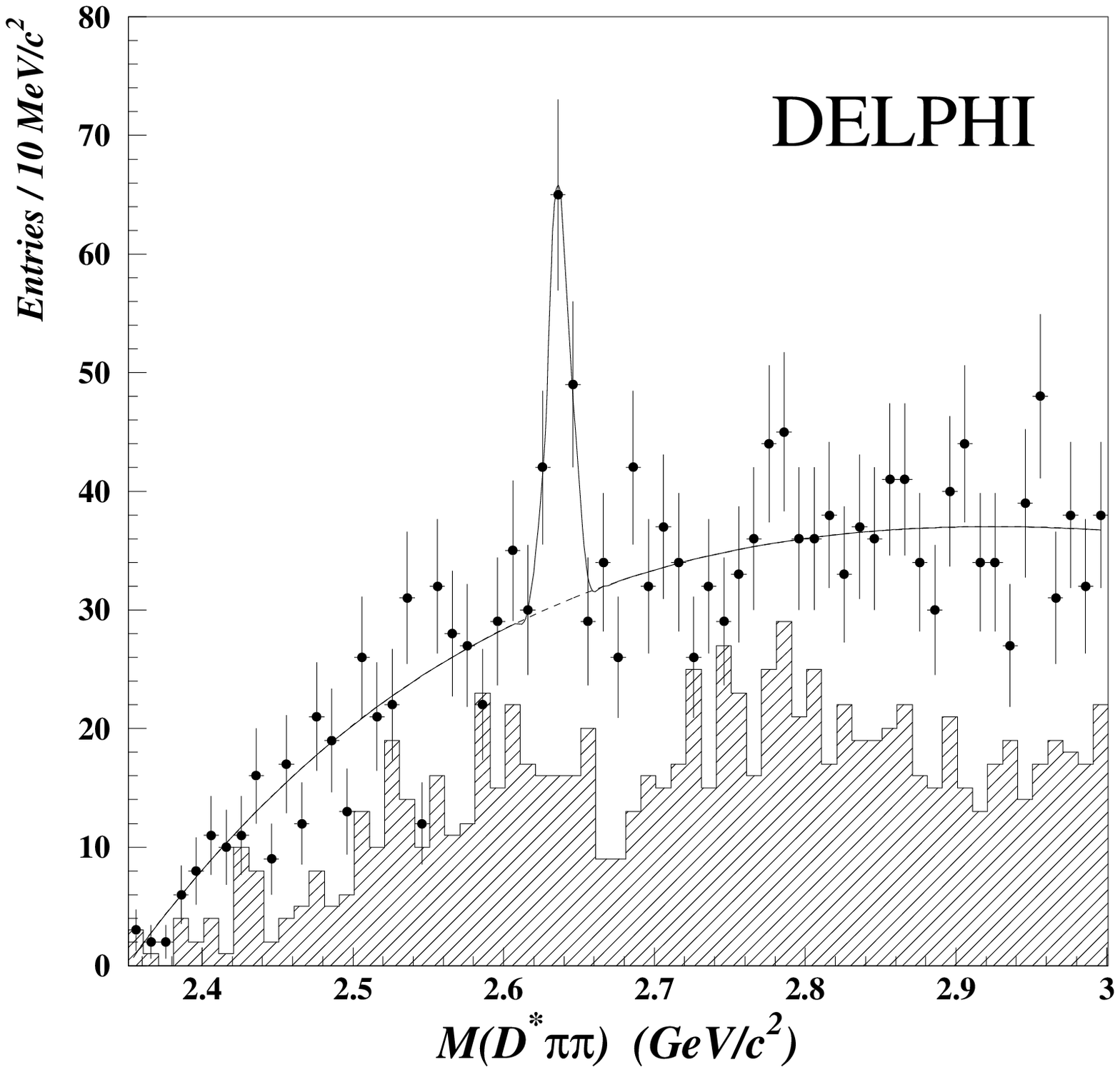,height=3.0in}\epsfig{file=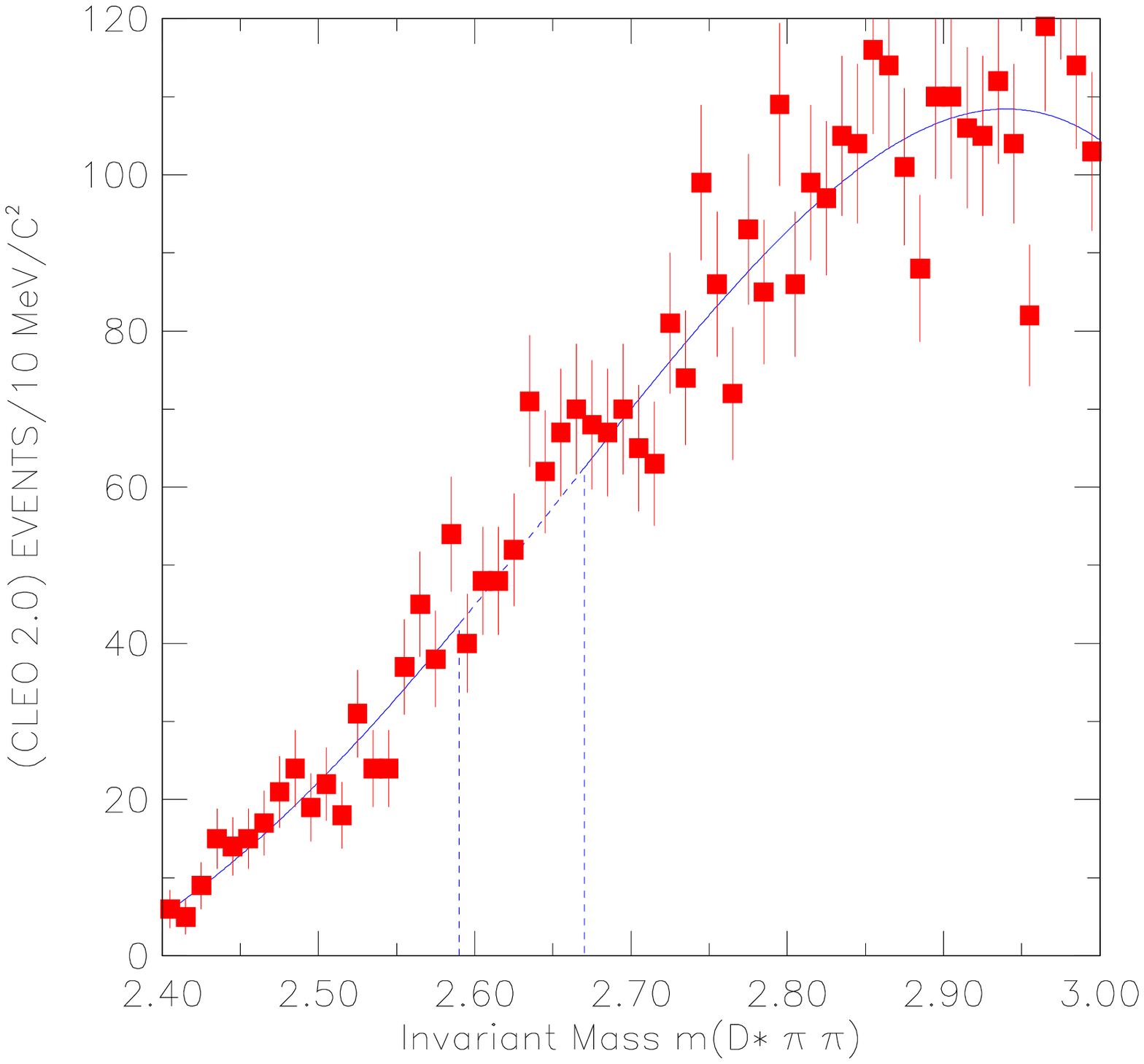,height=2.8in}}
\caption{The $D^{*+}\pi^-\pi^+$ invariant mass distribution from
DELPHI (left), from CLEO (right).}
\label{jrod:fig:dstrprime}   
\end{figure}

The analysis procedures used at CLEO are similar to those employed by
both DELPHI and OPAL. First pion and kaon candidates are selected from
tracks originating at the IP. These are then combined to form $D^0$
and $D^{*+}$ candidates requiring consistency with particle ID and
that the invariant masses be within the nominal values. To test the
reconstruction procedure and reduce the systematic errors, the $\dsp$
yields are compared to $D_J^0$ yields since the reconstruction
procedures differ only by a single charged pion. Also, the $\dsp$ to
$D^0_J$ production ratio allows for a more direct comparison between
the LEP and the CLEO results.

Using the entire \cleoii data set we have searched for the $\dsp$ in
the mass region suggested by the DELPHI results. We find no evidence
of an excess the region between 2590 MeV and 2670 MeV, and set a
preliminary upper limit of: 
\begin{equation}
\frac{N_{\dsp}\cdot {\cal B}(\dsp \rightarrow
D^{*+}\pi^-\pi^+)}{N_{D^{*0}_2}\cdot{\cal B}(D^{*0}_2\rightarrow
D^{*+}\pi^-)+N_{D^{0}_1}\cdot{\cal B}(D^{0}_1\rightarrow D^{*+}\pi^-)} < 0.16 \ @\ 90\% \ {\rm C.L.}
\label{jrod:eq:CLEOdsp}
\end{equation}
\noindent This may be compared with the DELPHI measurement of $0.49\pm
0.18\pm 0.10$ for this rate~\cite{jrod:DELPHI}. The invariant mass
distributions from DELPHI~\cite{jrod:DELPHI} and CLEO are shown in
\fig{jrod:fig:dstrpime}). The DELPHI result includes both $b\bar{b}$
and $c\bar{c}$ production.  

\subsection*{Measurement of \mbox{\boldmath $\Kpiws$ Decays}}
The decay $\Kpiws$ can proceed either a through doubly Cabbibo
suppressed decay (DCSD) channel or through $D^0-\bar{D}^0$ mixing. In
the standard model, the rate is expected at $0.3\%$ level so
this decay can be used to search for exotic or beyond-standard
model decay mechanisms. Standard model predictions for $R$, defined as 
$ R=\Gamma\left(D^0\rightarrow K^+\pi^-\right)/\Gamma\left(D^0\rightarrow K^-\pi^+\right)$, from mixing vary considerably from about $10^{-3}$ to $10^{-10}$. 
The contributions to $R$ from DCSD is of order $\tan^4\theta_C\sim
3\times 10^{-3}$\cite{jrod:DCSD93}. To separate the mixing and DCSD
contributions a measurement of the decay time distribution is
required. With the new silicon vertex detector, CLEO can now perform
this measurement, however, in the discussion that follows we
present only an $R$ measurement that includes contributions from
both mixing and DCSD. We plan to eventually add the time-dependent
measurement to this analysis.

\begin{figure}[t!] 
\centerline{\epsfig{file=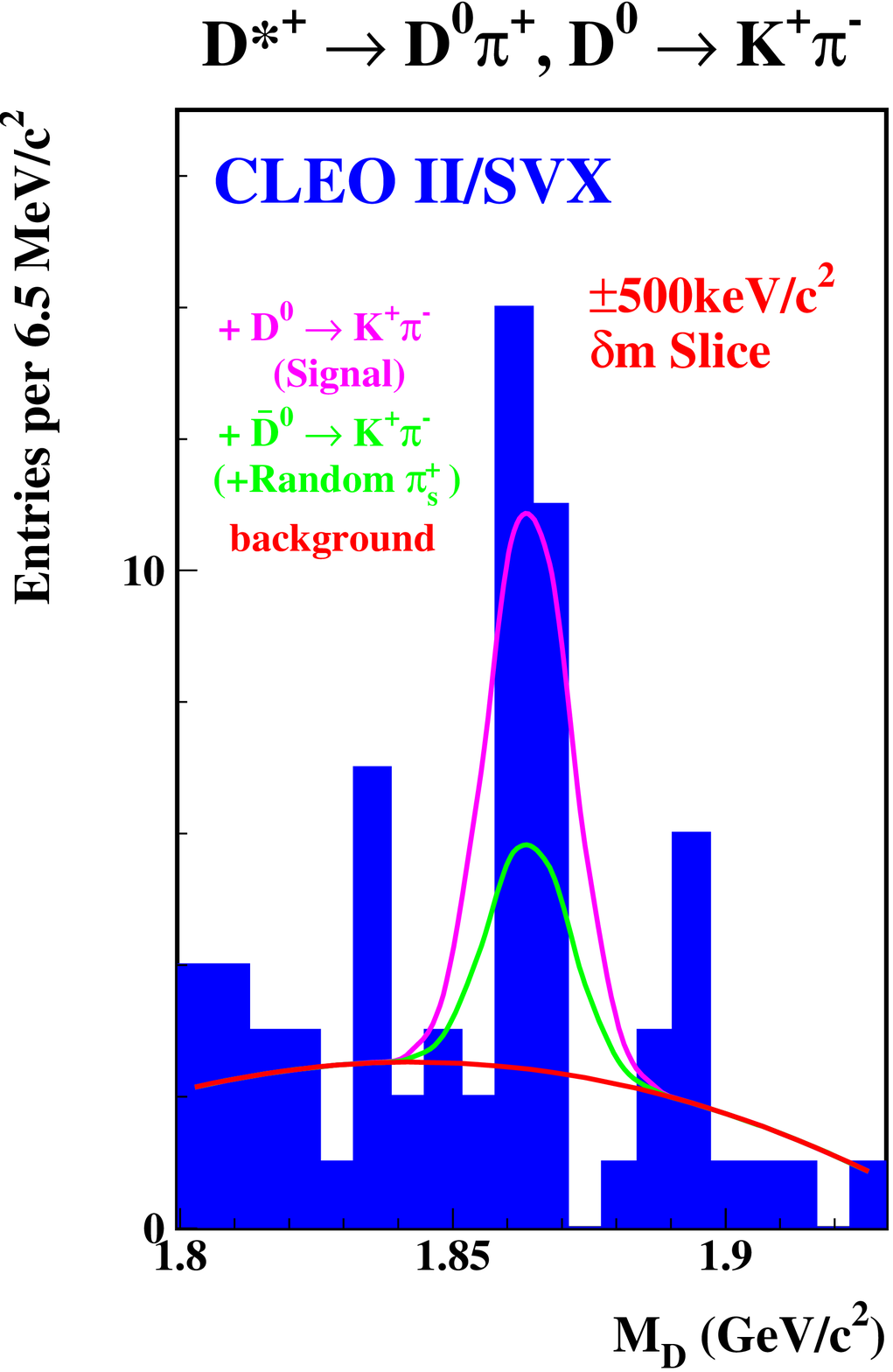,height=3.0in}\epsfig{file=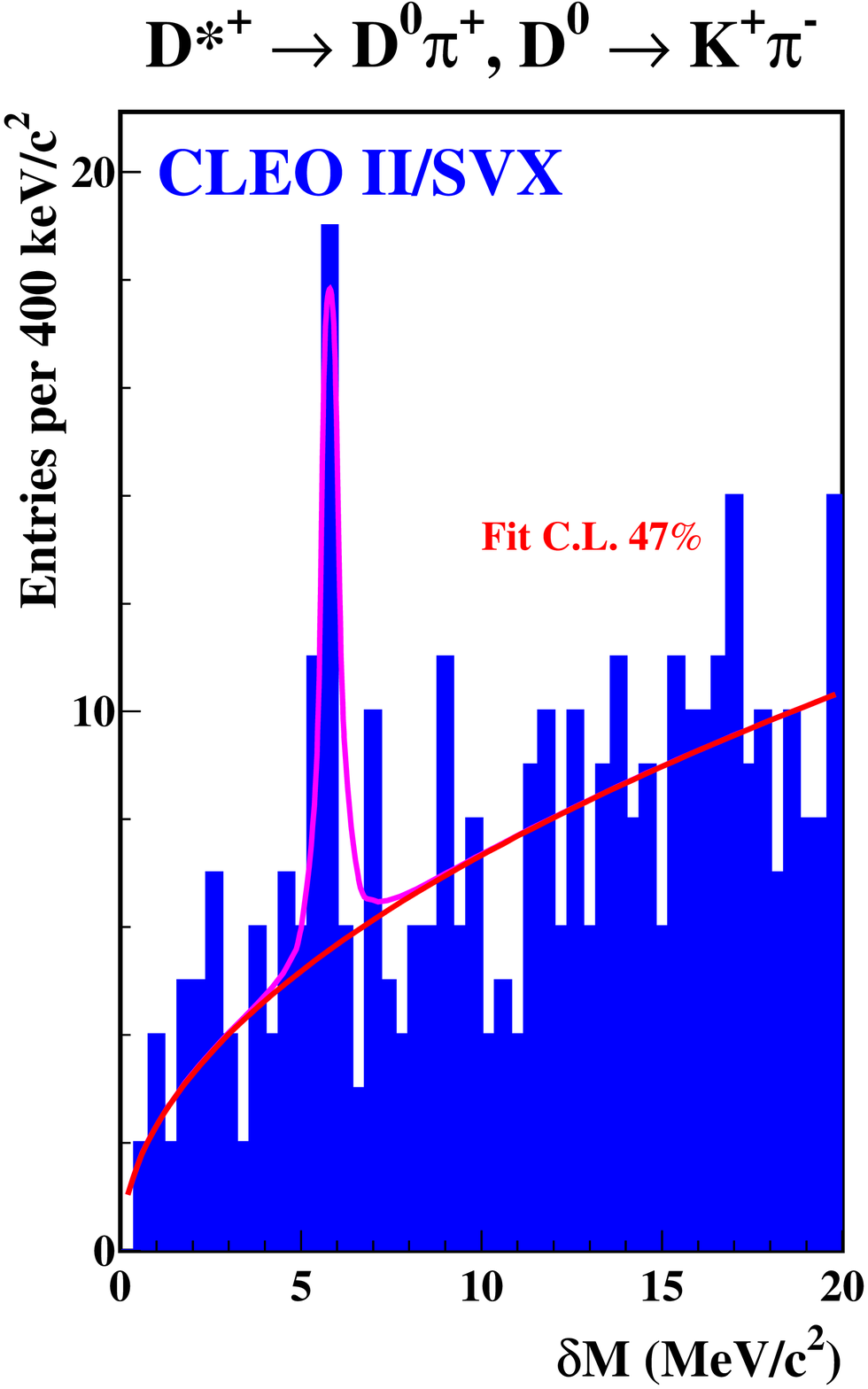,height=3.in}}
\caption{The wrong sign $M_{D^0}$ and $\delta M$ distributions with
the results of the 1D-ML fit with the signal and background
contributions superimposed on the data.}
\label{jrod:fig:D0toKpi}
 \end{figure}

To measure the ratio $R$ we have to determine the decay rates
$\Gamma(\Kpiws)$ and $\Gamma(\Kpirs)$. These rates are extracted from
analyzing high momentum continuum $D^{*+}\rightarrow D^0\pi^+_s
\rightarrow (K\pi)\pi^+_s$ events, where the sign of the slow pion
($\pi_s^+$) tags whether the $K\pi$ combination comes from a $D^0$ or
a $\bar{D}^0$. The combination where the sign of the $K$ and the slow
pion ($\pi^+_s$) are the same is referred to as the ``wrong sign''
combination while events where the $K$ and the $\pi_s^+$ have opposite
signs is referred to as the ``right sign'' combination. The
wrong/right sign signal yields are determined by fitting the
distribution of mass differences ($\delta M$) between the $D^{*+}$ 
and the $D^0$ and requiring the $D^0$ invariant mass ($M_D$) to be
within $\pm 13$ MeV ($2\sigma$) of the nominal $D^0$ mass. An
important feature in this 
analysis is the small width of both the $M_D$ and $\delta M$ mass
distributions as compared to other experiments. This is due primarily
to the improvements in the tracking algorithm and the vertex
resolution of the SVX detector. For example, in the \cleosvx data the
resolution of $M_D$ is now 6.5 MeV, while the $\delta M$ resolution is
200 keV compared with the \cleoii pre re-processed (data reconstructed
prior to the application of the Kalman algorithm) values of about 12
MeV and 750 keV respectively. The improved mass resolution allows for
a greater separation of signal from backgrounds. This improvement is
evident in the low levels of backgrounds in \fig~\ref{jrod:fig:D0toKpi}.

Because of the rarity of the $\Kpiws$ events, a significant amount of
work has been done to both reject and understand the backgrounds which
enter the $M_D$ and $\delta M$ distributions. The most significant
background components are due to real $\bar{D}^0\rightarrow K^+\pi^-$
combined with a random slow $\pi_s^+$ and backgrounds from
$D^{*+}\rightarrow (K^+\pi^-)\pi^+_s$ where the kaon and the pion are
mis-identified. The random slow $\pi^+_s$ background tends to peak in the
$D^0$ mass region but not in $\delta M$, while the doubly
mis-identified background tends to peak in $\delta M$ but not in the
$D^0$ mass. The latter is reduced by forming the $m_{\rm flp}(D^0)$,
where the mass assignments to the kaon and pion are switched, vetoing
the event if it's ``mass-flip'' mass is within $4\sigma$ of 
the nominal $D^0$ mass. The remaining backgrounds are modeled by Monte
Carlo simulations. The background distributions are used in the likelihood
function for the 1D maximum likelihood fit.

The preliminary result of the 1D ML fit for the ratio of decay rates
is:
\begin{equation}
\frac{\Gamma(\Kpiws)}{\Gamma(\Kpirs)} = 0.0032\pm 0.0012\pm 0.0015
\label{jrod:eq:wr-result}
\end{equation}
\noindent This result was obtained from an analysis of the current
\cleosvx dataset which includes both the off and on-resonance samples
for a total of 3.8 $fb^{-1}$. The new result is already more
statistically significant, by itself, than the current world average
of $0.0072\pm 0.0025$\cite{jrod:PDG}. There is currently about a
factor of two more \cleosvx data yet to be analyzed so we expect the
statistical significance to improve. We are also working on improving
the estimate of the systematic error and expected it to be much
smaller than the value quoted in \equ{\ref{jrod:eq:wr-result}}. Finally,     
it is worth noting that while compatible, within errors, with the
world average, the new $R$ measurement is lower by about a factor of 
two and therefore more consistent with theoretical expectations.

\subsection*{Summary and Conclusions}
We have presented preliminary results from five CLEO analysis in
hadronic decays of bottom and charmed mesons. From the $B$ hadronic
analyses we have shown results which provide us with the first
observation and measurements of the decay $\dstrdstr$, a first hint of
final state interactions in $B$ decays and a measurement of the decay
rates of the $B^-$ meson decaying into three of the $L=1$ charmed
mesons plus a single pion. From the $\ddblstrpi$ analysis we have the
first observation of the broad $L=1,j_l=1/2$ charmed meson and have
determined its mass and decay width. Our preliminary search for the
first radial excitation of the charmed meson (the $\dsp$ ) has been
unable to confirm the observation by the DELPHI collaboration. Finally, 
our preliminary results on a measurement of the ratio
$\Gamma(\Kpiws)/\Gamma(\Kpirs)$ gives a results lower than previous
measurements. This new measurement is an improvement over the previous
CLEO results with a substantial increase in data and improved
detector.

\end{document}